# A Regularized Spatial Market Segmentation Method with Dirichlet Process Gaussian Mixture Prior

Won Chang

*Department of Mathematical Sciences, University of Cincinnati, Cincinnati, Ohio, USA*

E-mail: Won.Chang@uc.edu

Sunghoon Kim

*W.P. Carey School of Business, Arizona State University, Tempe, Arizona, USA*

Heewon Chae

*W.P. Carey School of Business, Arizona State University, Tempe, Arizona, USA*

**Summary.** Spatially referenced data are increasingly available thanks to the development of modern GPS technology. They also provide rich opportunities for spatial analytics in the field of marketing science. Our main interest is to propose a new efficient statistical framework to conduct spatial segmentation analysis for restaurants located in a metropolitan area in the U.S. The spatial segmentation problem poses important statistical challenges: selecting the optimal number of underlying structures of market segments, capturing complex and flexible spatial structures, and resolving any possible small n and large p issue which can be typical in latent class analysis. Existing approaches try to tackle these issues in heuristic ways or seem silent on them. To overcome these challenges, we propose a new statistical framework based on regularized Bayesian spatial mixture regressions with Dirichlet process integrating ridge or lasso regularization. Our simulation study demonstrates that the proposed models successfully recover the underlying spatial clustering structures and outperforms two existing benchmark models. In the empirical analysis using online customer satisfaction data from the Yelp, our models provides interesting insights on segment-level key drivers of customer satisfaction and interpretable relationships between regional demographics and restaurants' characteristics.

# 1. Introduction

The modern communication and network technologies (e.g., smartphones, PC or IoT (Internet of Things)) produce an explosive increase in various types of automatically and dynamically generated data (e.g., user generated data, online click stream data, etc.), which is often referred to as 'Big Data phenomenon'. The large availability of various consumer data derives recent needs and popularity of business analytics (see Fader and Winer 2012). Notably, many of recent data are accompanied by geographic information thanks to geographic systems such as GPS and built-in location softwares/apps. This provides rich opportunities for spatial analysis in business as managers may be able to utilize the geographic information (e.g., spatial market segmentation, customer relationship management across geographic areas, etc.) to create and sustain competitive advantages (Oracle 2010, Shimonti 2018).,However, spatial models have been relatively understudied in the field of marketing while the current methodological development in spatial statistics has mainly focused on environmental sciences (see Bradlow et al. 2005, Bronnenberg 2005).

Market segmentation is a first and important step for managers to set up an adequate marketing strategy by understanding potential heterogeneity amongst consumers. Given that academics and practitioners rank market segmentation as having the strongest impact in the field of marketing (Roberts, Kayande, and Stremersch 2014), a stream of statistical models has been developed and applied to dealing with segmentation problems (e.g., Wedel and Kamakura 2000, Ter Hofstede et al. 2002, Kim, Fong and DeSarbo 2012, Govind et al. 2018). With the recent large availability of spatial information, our goal is to develop a better spatial segmentation method to help managers to develop more sophisticated and effective regional strategies. Spatial segmentation of retail businesses (e.g., restaurants, retailers, and hotels), which are strongly influenced by the characteristics of local communities, may exhibit some underlying geographic structures, such that nearby locations with spatial dependency can probabilistically fall into a same market segment. When an underlying geographic structure is present, reflecting the geographic information in a modeling framework can significantly improve the market segmentation performance (see Ter Hofstede et al. 2002, Govind et al. 2018).

In this paper, we focus on spatial segmentation analysis of restaurants using user generated review data from the Yelp. The Yelp data provide a good source to explore market segmentation problems in business by offering a number of characteristics of retail businesses with their spatial information (e.g., longitudes and latitudes of retail establishments). For example, Chae (2016), who also utilized the Yelp data, finds that consumers' socio-economic demographics in a local area influence the way restaurants located in the region pursue their menu diversification strategies (i.e. how many culinary categories they include in the menus). The author conjectured that the strategies chosen by the restaurants reflect local consumers' preferences, which are heavily affected by their socio-economic status. Chae (2016)'s study, however, did not investigate a potential underlying spatial heterogeneity structure (i.e., spatial segments) nor possible spatial dependency.

Spatial segmentation with our Yelp dataset poses some important statistical challenges. First, the optimal number of market segments needs to be parsimoniously and efficiently estimated based on the observed data. Whereas most existing methods in spatial segmentation (e.g., Hofstede et al., 2002, Govind et al., 2018) have used heuristics ways (e.g., BIC), our proposed procedure automatically selects the optimal number of segments by adopting the Dirichlet stochastic process. Second, the spatial segments tend to have highly complicated shapes because they are likely to be determined by various features. For example, multiple regions with similar characteristics can be found in the opposite sides of a metropolitan area, as they share similar characteristics such as similar distances from the center of the city and main property types (e.g. commercial area vs. office area). Relatively simple segmentation approaches reflecting spatial dependence such as a Markov random field-type model (see Hofstede et al. 2002) are not flexible enough to accommodate such complex spatial structure. Third, this type of segmentation models would be vulnerable to the sample size issue because the clustering models may (theoretically) have segments containing a very small number of members. Additionally, considering recent challenges of high-dimensional data (i.e., many independent variables in regression), the sample size issue can be aggravated for the segmentation models. Thus, a certain form of regularization (i.e., feature selection) should be applied for sound parameter estimation in spatial segmentation procedures, and addressing this issue also involves answering the question of which regularization scheme (e.g., lasso vs. ridge) is preferable.

Our aim is to develop a better statistical method that addresses these issues for the spatial segmentation (i.e., strategic segmentation of restaurants in the Washington DC metropolitan area). We develop a Bayesian approach that combines a spatial Dirichlet process and the mixture regression framework, which can accommodate both the ridge and lasso regularization under the unified prior scheme. Given that the three pre-described challenges are commonly faced in market segmentation analysis, our methodological development can be broadly applied to other similar segmentation problems beyond the domain of business. In the following Section 2, we describe data and variables in our Yelp dataset.

## 2. Data Description

Our main focus is to conduct spatial segmentation analysis with customer satisfaction data for the restaurants located in the greater Washington DC metropolitan area, composed of the District of Columbia and municipalities in 22 counties in Virginia, Maryland, and West Virginia. The sample data include 2,065 restaurants in 61 cities with 155 zip codes that have more than 15 consumer reviews. Information on the sample restaurants was collected from Yelp, a major online source of information on local businesses. All the information on restaurants was downloaded on February 27, 2014.

*Dependent variable*. It is measured as an average rating by customers who posted their reviews with a 5-point scale, ranging from 1 ("completely dissatisfied") to 5 ("completely satisfied") in 2014. Because restaurant characteristics to be included as explanatory variables, described below, are the attributes in 2014 at the time of extracting the data and they may have been changed in the past, we include ratings posted only in 2014 to accurately estimate the influence of restaurant characteristics on customer satisfaction.

*Independent variables*. Four types of restaurant characteristics are included as potentially regionally heterogeneous predictors of customer satisfaction—category spanning (Hannan, Polos, and Carroll 2007; Rao, Monin, and Durand 2005), price level, chain yes/no, and fast food yes/no. First, we included a variable measuring restaurant's degree of "cuisine category spanning" to understand regionally heterogeneous relationship between the level of category spanning and customer satisfaction. *Category Spanning* (CSPAN) is defined as the extent to which a business organization (e.g., a restaurant in this paper) has membership in multiple market (i.e., cuisine) categories (Hannan et al. 2007; Rao et al. 2005). For example, whereas some restaurants claim membership in a single culinary category (e.g., American), others span multiple categories and offer dishes from various culinary origins (e.g., American, Mexican, and Chinese). A recent stream of research in business shows that firms' CSPAN strategy is closely related to the performance of restaurants and such effect varies depending on restaurants' and customers' characteristics (Chae 2016; Kovács and Johnson 2014). Also, the CSPAN strategy of restaurants may be evaluated differently across regions where customers have different preferences depending on their socio-economic status. All establishments listed in the restaurant group on Yelp are categorized into one or more of 108 cuisine styles, such as American, French, Korean, and so on. About 42% of the sample belong to one cuisine style, 35% of the restaurants are in two styles, and the rest 23% belong to three or more cuisine styles.

The category spanning is measured by using:

$$\text{Category Spanning (CSPAN)} = \begin{cases} 0 & \text{if } numcate = 1; \\ numcate \times \bar{d}x & \text{if } numcate > 1 \end{cases}, \quad (1)$$

where *numcate* denotes the number of categories the organization claims and $\bar{d}x$ denotes the average cosine distance between the categories. The cosine distance between each pair of categories on a co-occurrence matrix shows which categories are claimed together by restaurants. In the Equation (1), we took into account the distance between the categories a restaurant spans when measuring the degree of spanning following previous literature because some cuisine styles share more attributes and are considered closer than others (Chae 2016; Kovács and Hannan 2010). Including cosine distance between categories allows the similarity structure of the categories spanned to be captured (Leahey, Beckman, and Stanko 2016; Sohn 2001). For example, if the cosine distance between Indian and Pakistan is 0.23 and that between Indian and Italian is 0.99, the CSPAN score of a restaurant that spans Indian and Pakistan categories is 0.46 (=2×0.23) while the score of a restaurant that spans Indian and Italian is 1.98 (=2×0.99).

As the second explanatory variable, the "price level" can be related to customer satisfaction and is included with a scale of 1 to 4. Yelp uses four categorical system to classify the price level of restaurants, indicating "the approximate cost per person for a meal, including one drink, tax, and tips": $ denotes "under US$10," $$ denotes "US$11–US$30," $$$ denotes "US$30–US$61," and $$$$ denotes "above US$61." In our data, 22% of the restaurants are in the $ (lowest price) range, 66% are in the $$ range, 10% are in the $$$ range, and the remaining 2% are in the $$$$ (highest price) range. As the third and fourth variables, we include two dummy variables for "chain restaurants" (1 for yes; 0 for no) and "fast food restaurants" (1 for yes; 0 for no), which can capture their unique business characteristics (Sloane, Lewis, and Nascimento 2005).

## 3. Existing Spatial Mixture Regressions

### 3.1. Basic Framework

We describe the basic framework of spatial mixture regressions for analyzing the customer satisfaction data introduced in the previous section. We let $n$ be the number of observations in the dataset and $p$ be the number of independent variables (i.e. the number of features used to predict the mean customer satisfaction rating). For each rescaled spatial location $s_i \in D = [0,1] \times [0,1]$ ($i = 1, \dots, n$), we let the scaler $y_i$ denote the response variable (e.g., mean rating) and the $1 \times p$ vector $\mathbf{x}_i$ denote the independent variables (features). We assume that the observed data $\{y_i, \mathbf{x}_i, s_i\}_{i=1}^n$ form meaningful segments depending both on *the spatial* locations $s_1, \ldots, s_n$ and *the input-output relationship*, i.e., the relationship between the response ($y_i$) and independent variables ($\mathbf{x}_i$, a $1 \times p$ vector). We denote the membership of the $i$th observation for these unobserved segments by $g_i \in \{1, \ldots, K^*\}$, where $K^*$ is the number of latent segments.

At each spatial location $s_i$, we model the association between (restaurant) features and customer satisfaction, where the form of association is determined by the segment membership $g_i$. This can be represented as a regression model that relates the features $\mathbf{x}_i$ to the mean rating $y_i$ through the following cluster dependent linear regression model:

$$y_i = \mathbf{x}_i \boldsymbol{\beta}_{g_i} + \epsilon_i. \tag{2}$$

The cluster dependent regression coefficient $\boldsymbol{\beta}_{g_i} \in \{\boldsymbol{\beta}_1, \boldsymbol{\beta}_2, \ldots, \boldsymbol{\beta}_{K^*}\}$ ($p \times 1$ vector) defines how the features affect the mean rating depending on the membership of the $i$th observation $g_i \in \{1, \ldots, K^*\}$. Given the cluster specific variance parameter $\sigma_{g_i}^2 \in \{\sigma_1^2, \sigma_2^2, \ldots, \sigma_{K^*}^2\}$, we assume that the variance of the error term $\epsilon_i$ is inversely proportional to the number of ratings for the $i$th observation $N_i$ (i.e., $\epsilon_i \sim N(0, \sigma_{g_i}^2/N_i)$), to reflect the fact that $y_i$ is the mean of $N_i$ customer ratings. This configuration leads to an inference procedure based on the weighted least squares estimation (see Section 4.3 below). In this formulation, the ratings of restaurants that received more number of customer ratings have more impact on determining the posterior densities of the coefficients $\boldsymbol{\beta}_1, \boldsymbol{\beta}_2, \ldots, \boldsymbol{\beta}_{K^*}$.

### *3.2. Challenges in Market Segmentation Analysis*

Applying the statistical framework described in Section 3.1 to the data introduced in Section 2 poses the following important challenges: First, the number of true cluster memberships $K^*$ is not known a priori and hence needs to be estimated based on observed data. The existing approaches in Marketing, the Spatial-Association Model (SAM) proposed by Hofstede et al. (2002) and the Spatially Dependent Segmentation (SDS) by Govind et al. (2018), use heuristic model selection approaches (e.g., Bayes factor, BIC), which often perform poorly when applied to spatial clustering as shown by our simulation study in Section 5 below. Second, each spatial segment potentially has a highly complicated shapes and even consists of multiple separate regions, because neighborhoods with similar characteristics can be found in a number of separate locations. Third, inference on the regression coefficients in the model in Equation (2) may require some form of regularization. For each segment $g$, the number of observations $n_g = \sum_{i=1}^{n} I(g_i = g)$ may not be large enough to reliably estimate $p$ number of parameters in $\boldsymbol{\beta}_g$. Such phenomenon can be more common in the modern marketing application as the firms are increasing the number of product/service attributes to differentiate their products/services (see Kim, DeSarbo, Fong 2018) and consumers' choice tend to become diversified, indicating that managers need to manage many smaller segments. This issue can be exacerbated when MCMC is used for inference, as the chain will search through a wide range of possible label configuration and frequently visit states where at least one cluster is empty or has only few observations in it. Even if the true number of observations in each cluster is enough, some form of regularization is necessary to make sure that MCMC chain can discover the optimal labels. Any similar search algorithm to find the cluster labels for observations will encounter very similar issues.

### *3.3. Previous Spatial Segmentation Methods in Marketing*

Hofstede et al. (2002) pioneered spatial mixture regression modeling in Marketing, in the context of regression analysis of areal data, by incorporating spatial clustering in segment memberships through the dependence between neighboring areas. Their fully Bayesian approach uses Bayes factors to determine the number of clusters and it is not straightforward how to adapt their approach to our point-referenced dataset. Another existing approach, recently introduced by Govind et al (2018), is based on geographically weighted regression (GWR). As a first limitation, unlike the EM algorithm, their inference procedure does not enjoy the theoretical optimality because the assumed likelihood functions for their coefficient estimation and membership estimation do not coincide. Second, the likelihood function used in their BIC computation for estimating the true number of clusters ($K^*$) do not match the assumed likelihood function for coefficient estimation. Thus, it is difficult to establish the theoretical properties of the inference results from this framework.

## 4. The Regularized Bayesian Spatial Mixture Regression with Dirichlet Process Prior

We now develop a Dirichlet process Gaussian mixture-based spatial regression framework with regularization on regression coefficients. The proposed framework has been developed with the following features: (i) automated estimation of the number of latent segments, (ii) capturing highly flexible shape of spatial segments, and (iii) regularized regression per each derived segment for stable inference by preventing any potential large *p* and small *n* issues. We call the proposed method as regularized Bayesian spatial mixture regression with Dirichlet process prior (RSD) thereafter. We introduce each component of the RSD.

### 4.1. Dirichlet Process Prior for Mixture Probabilities of Spatial Segments

To automatically choose the number of spatial clusters, we set up Dirichlet process prior for the segment membership distribution. As a finite approximation to the infinite dimensional Dirichlet process (Ishwaran and James, 2001), we first assume that the distribution of the individual cluster membership has a categorical distribution with class probabilities $q_1, q_2, \ldots, q_K$:

$$g_i \sim \text{Categorical}(q_1, q_2, \ldots, q_K),$$

with the following stick breaking prior (Sethuraman, 1994):

$$q_g = \begin{cases} U_1, \text{ for } g = 1 \\ U_g \prod_{k=1}^{g-1}(1 - U_k), \text{ for } g = 2, \ldots, K \end{cases} \quad (3)$$

where $U_1, \ldots, U_K \sim iid\ Beta(1, b_U)$ with a rate parameter $b_U$. For $g = 2, \ldots, K$, $\prod_{k=1}^{g-1}(1 - U_k)$ can be rewritten as $1 - \sum_{k=1}^{g-1} q_k$, which is the 'remaining probability' that has not been accounted for the leading $g - 1$ mixture components. Each beta random variable $U_g$ represents the proportion assigned to the $g^{th}$ segment, out of the remaining probability $1 - \sum_{k=1}^{g-1} q_k$. Theoretically, the stick breaking prior assumes infinite number of mixture components $(K = \infty)$, but practically one can approximate its distribution using a reasonably large finite $K$. The remaining probability $1 - \sum_{k=1}^{g-1} q_k$ for a large $g$ becomes negligibly small, and therefore many mixture components become empty; i.e., contain no observations in them. Through this mechanism this finite approximation to the stick-breaking model can choose the number of segments $K^*$ in an automated fashion, which provides a clear methodological advantage over other existing clustering approaches using model selection heuristics (e.g., Bayes factors or BIC). We set a non-informative prior for $b_U$ given by $b_U \sim \text{Gamma}(0.1, 0.1)$.

## *4.2. Spatial Segmentation Using Dirichlet Process Gaussian Mixture*

To allow for highly flexible shapes of spatial segments, we use the spatial Dirichlet Process Gaussian mixture model within each segment (Reich and Bondell, 2011). We model the locations within a latent segment, denoted by $g \in \{1, \ldots, K\}$, using the following Gaussian mixture density with $M$ components:

$$f_g(\boldsymbol{s}_i) = \sum_{h=1}^{M} p_{gh} N(\boldsymbol{s}_i | \boldsymbol{\mu}_{gh}, \Sigma_g), \quad (4)$$

where $0 \leq p_{g1}, \ldots, p_{gM} \leq 1$ are the mixing proportions, and $N(\boldsymbol{s}_i|\boldsymbol{\mu}_{gh}, \Sigma_g)$ is the Gaussian density function with 2-dimensional mean vector $\boldsymbol{\mu}_{gh}$ and 2-by-2 covariance matrix $\Sigma_g$. The subscript $h$ represents the membership for the $M$ possible different mixture components of the segment $g$. By allowing each segment $g$ to have multiple sub-components, we can create a highly flexible shape of clusters for each $g$.

We set $\Sigma_g = \tau_g^2 I_2$ with 2 by 2 identity matrix $I_2$ and the variance parameter $\tau_g^2 > 0$. This specification leads to a model with better identifiability without sacrificing flexibility much. The component mean $\boldsymbol{\mu}_{gh}$ and precision $\tau_g^2$ receive the following vague conjugate hyperpriors:

$$\boldsymbol{\mu}_{gh} \sim \text{TBN}(\mathbf{0}, \tau_0^{-2}, -1, +1),$$

$$\tau_g^2 \sim \text{Gamma}(a_\tau, b_\tau),$$

with $\tau_0^{-2} = 1000$, $a_\tau = 0.1$, and $b_\tau = 0.1$. Note that $\text{Gamma}(a, b)$ denotes a gamma distribution with shape parameter $a$ and scale parameter $b$ and $\text{TBN}(\mathbf{m}, v, -1, 1)$ denotes a truncated bivariate normal distribution with the following probability density:

$$f(x, y) \propto \phi\left(\frac{x - m_1}{v}\right) \phi\left(\frac{y - m_2}{v}\right) I(-1 < x < 1) I(-1 < y < 1),$$

where $\phi$ is the standard normal probability density; $v > 0$ is the variance, and $m_1$ and $m_2$ are the first and second elements of the 2×1 mean vector $\mathbf{m}$.

Using additional membership variables $h_1, \ldots, h_n \in \{1, \ldots, M\}$ for the mixture components, one can represent this Gaussian mixture model in the following equivalent form:

$$\boldsymbol{s}_i | g_i = g, h_i = h \sim N(\boldsymbol{\mu}_{gh}, \tau_g^{-2} I_2)$$

$$h_i | g_i = g \sim \text{Categorical}(p_{g1}, \ldots, p_{gM}).$$

Therefore, the likelihood that each location $s_i$ belongs to the mixture component $h$ of the segment $g$ quickly increases as the distance between $s_i$ and $\mu_{gh}$ decreases, in the order of $\exp(-\tau_g^2 d^2)$, where $d$ is the Euclidean distance between the two points. This model configuration encourages nearby locations to fall into the same cluster, because any close locations will be also close to the same cluster center $\mu_{gh}$.

In a similar fashion to the prior specification for membership probabilities $q_1, \ldots, q_K$, the mixing proportions $p_{g1}, \ldots, p_{gM}$ receive the following stick-breaking prior;

$$p_{gh} = \begin{cases} V_{g1}, & for\ h = 1 \\ V_{gh} \prod_{k=1}^{h-1}(1 - V_{gk}), & for\ h = 2, \ldots, M \end{cases} \quad (5)$$

where $V_{g1}, \ldots, V_{gM} \sim iid\ Beta(1, b_V)$ with a rate parameter $b_V$. This prior configuration, in combination with the mixture model in (4), leads to a Dirichlet process Gaussian mixture model (Antoniak, 1974). A smaller value of $b_V$ leads to a flatter prior for the number of non-empty clusters. Following Reich and Bondell (2011), we use a hyperprior given by $b_V \sim Gamma(1, 1/4)$ to encourage the number of non-empty clusters take a value between 1 and 15.

Based on the mixture model specification above, the cluster membership for each observation can be sampled from

$$g_i | all\ others \sim Categorical(\tilde{q}_{i1}, \ldots, \tilde{q}_{iK}),$$

where the probability for each category $\tilde{q}_{ig}$ is given by

$$\tilde{q}_{ig} | all\ others \propto q_g \frac{1}{\sigma_g/\sqrt{N_i}} \exp\left(-\frac{(y_i - \mathbf{x}_i \boldsymbol{\beta}_g)^2}{\sigma_g^2/N_i}\right) \cdot \sum_{h=1}^{M} p_{gh} \tau_g \exp\left(-\tau_g^2 \|s_i - \mu_{gh}\|^2\right), \quad g = 1, \ldots, K,$$

where $\|.\|$ is the norm of a vector. Therefore, the cluster member probabilities are determined by both the relationship between $y_i$ and $\mathbf{x}_i$ (represented by the first term) and the geographical clustering (represented by the second term). The probability $q_g$ given by the stick breaking prior explained above in Section 3.2 encourages that only a part of $K$ different clusters are non-empty. Once the cluster membership is sampled, we also sample the component membership from

$$h_i|\text{all others} \sim \text{Categorical}(\tilde{p}_{i1}, \ldots, \tilde{p}_{iM}),$$

where the probability for each component $\tilde{p}_{ih}$ ($h = 1, \ldots, M$) is given by

$$\tilde{p}_{ih}|\text{all others} \propto p_{g_ih}\tau_{g_i}^2 \exp\left(-\tau_{g_i}^2\|\mathbf{s}_i - \boldsymbol{\mu}_{g_ih}\|^2\right).$$

Again, the probability $p_{g_ih}$ from the stick breaking prior in Equation (5) encourages that only a part of the $M$ components are non-empty.

The full conditional distribution for each mixture component mean $\boldsymbol{\mu}_{gh}$ is given by the following truncated bivariate normal distribution:

$$\boldsymbol{\mu}_{gh}|\text{all others} \sim \text{TBN}(\mathbf{m}_{gh}, v_{gh}^{-2}\mathbf{I}_2, -1, +1),$$

with $\mathbf{m}_{gh} = \frac{\tau_g^2 \sum_{i=1}^n \mathbf{s}_i I(g_i=g) I(h_i=h)}{\tau_0^2 + \tau_g^2 \sum_{i=1}^n I(g_i=g) I(h_i=h)}$, $v_{gh}^2 = \tau_0^2 + \tau_g^2 \sum_{i=1}^n I(g_i = g) I(h_i = h)$, and $I(.)$ is an indicator function which takes the value of 1 when the condition in $(.)$ is true and 0 otherwise. We sample the mixture component precision $\tau_g^2$ from the following conjugate posterior:

$$\tau_g^2|\text{all others} \sim \text{Gamma}(\tilde{a}_{\tau,g}, \tilde{b}_{\tau,g}),$$

where $\tilde{a}_{\tau,g} = a_\tau + \sum_{i=1}^n I(g_i = g)$ and $\tilde{b}_{\tau,g} = b_\tau + \frac{\sum_{i=1}^n \|\mathbf{s}_i - \boldsymbol{\mu}_{g_ih}\|^2 I(g_i=g)}{2}$.

### 4.3. Regularized Estimation for Regression Coefficients

As pointed out in Section 3.2, inference on the regression coefficients $\boldsymbol{\beta}_1, \boldsymbol{\beta}_2, \ldots, \boldsymbol{\beta}_K$ should be regularized through some priors on them as the number of observations $n_g$ in at least one segment can be small. We consider the following two different choices, each of which provides different degrees of regularization in inference on the coefficients:

1) The coefficients $\boldsymbol{\beta}_g$ ($g = 1, \ldots, K$) for each segment are given a vague conjugate prior $N(\mathbf{0}, \sigma_g^2 \Lambda_0)$ with $\Lambda_0 = c\mathbf{I}_p$, where $\mathbf{I}_p$ is the $p$-by-$p$ identify matrix and $c$ is a very small constant. We thereafter set $c$ to 0.01, but model fitting result is largely unaffected by the choice for $c$ as long as it is restricted to a small value. This standard conjugate specification corresponds to the ridge regression and imposes a weak regularization as long as $c$ is small. We call this specification the 'ridge prior' thereafter.

2) The coefficients receive a Laplace prior, defined as

$$f(\boldsymbol{\beta}_g | \sigma_g^2) = \prod_{j=1}^{p} \frac{\lambda}{2\sigma_g} \exp\left(-\frac{\lambda |\beta_{gj}|}{\sigma_g}\right), g = 1, \ldots, K \qquad (6)$$

where $\lambda$ determines the strength of regularization. This prior specification corresponds to the lasso constraint, which tend to impose a stronger regularization than the ridge regression. We set $\lambda = 0.03$ to balance between prediction performance and interpretability in the simulation study (Section 5) and real data analysis (Section 6). While this specification leads to a non-conjugate posterior for $\boldsymbol{\beta}_g$, sampling from the resulting posterior density can be easily done by a series of Gibbs sampling steps (Park and Casella, 2008, see Section 3.3 for details). We call this prior as the 'lasso prior' thereafter. To complete the model specification, we assume that the variance parameters $\sigma_1^2, \ldots, \sigma_K^2$ receives a common vague conjugate prior InvGamma($a_\sigma, b_\sigma$), where InvGamma($a, b$) denotes an inverse gamma distribution with shape and scale parameters $a$ and $b$.

The full conditional distributions of the regression coefficients $\boldsymbol{\beta}_g$'s and the error variances $\sigma_g^2$'s for the model in (2) are given by

$$\boldsymbol{\beta}_g | \text{all others} \sim N\left(\left(\mathbf{X}_g^T \mathbf{W}_g \mathbf{X}_g + \mathbf{D}_g^{-1}\right)^{-1} \mathbf{X}_g^T \mathbf{W}_g \mathbf{Y}_g, \sigma_g^2 \left(\mathbf{X}_g^T \mathbf{W}_g \mathbf{X}_g + \mathbf{D}_g^{-1}\right)^{-1}\right), \qquad (7)$$

$$\sigma_g^2 | \text{all others} \sim \text{InvGamma}(\tilde{a}_{\sigma,g}, \tilde{b}_{\sigma,g}),$$

where $X_g$ ($n_g$-by-$p$ matrix with $n_g = \sum_{i=1}^{n} I(g_i = g)$) and $Y_g$ ($n_g$-by-1 vector) are the collection of $\mathbf{x}_i$'s and $Y_i$'s for all $i$'s satisfying $g_i = g$; $W_g$ is an $n_g \times n_g$ diagonal matrix whose diagonal elements are $N_i$'s corresponding to the elements in $Y_g$; $\tilde{a}_{\sigma,g} = a_\sigma + \frac{n_g}{2}$ and $\tilde{b}_{\sigma,g} = b_\sigma + \frac{\sum_{i=1}^{n} N_i(y_i - \mathbf{x}_i \boldsymbol{\beta}_g)^2 I(g_i=g)}{2}$; $D_g = \mathrm{diag}(\psi_{g1}, \ldots, \psi_{gp})$ is a $p$-by-$p$ matrix whose diagonal elements are determined by the prior variances for $\boldsymbol{\beta}_g$. Note that both the ridge and lasso variable selection priors can be specified under the proposed framework: For the ridge prior defined above, we can simply set $\psi_{g1} = \cdots = \psi_{gp} = 1/c = 100$. For the lasso prior, we sample $\psi_{g1}, \ldots, \psi_{gp}$ based on the following conditional distributions (Park and Casella, 2008):

$$1/\psi_{gj} | \text{all others} \sim \mathrm{InvGaussian}(\mu'_{gj}, \lambda^2)$$

where $\mu'_{gj} = \sqrt{\frac{\lambda^2 \sigma_g^2}{\beta_{gj}^2}}$, $D_g = \mathrm{diag}(\psi_{g1}, \ldots, \psi_{gp})$ and $\mathrm{InvGaussian}(a, b)$ denotes the inverse-Gaussian distribution with the following probability density (Chhikara and Folks 1989):

$$f(a,b) = \sqrt{\frac{b}{2\pi}} x^{-3/2} \exp\left(-\frac{b(x-a)^2}{2a^2 x}\right), x > 0, -\infty < a < \infty, b > 0.$$

Park and Casella (2008) showed that the resulting marginal density for $\boldsymbol{\beta}_g$ from this sampling scheme is the same as the posterior density given by the Laplace density in Equation (6).

We expect that these two aforementioned priors (e.g., ridge prior and lasso prior) will lead to different selection of restaurants features. Selecting key features has important implications in Business as it informs how managers can efficiently and effectively allocate their limited resource to improve the customer satisfaction. We expect that the stronger regularization imposed by the lasso regularization results in more sparsity (i.e. more stronger shrinkage towards zero), strongly ruling out irrelevant features for consumer satisfaction compared to the ridge regularization. However, in some cases, the lasso regularization might cause too much shrinkage, especially when the number of predictors $p$ is much less than the number of observations $n$, resulting in less efficient prediction and parameter estimation than the ridge regularization. Note, the positive effect of lasso scheme is salient in the cases of $p$ is larger than $n$ which are popular in biostatistics. Thus, it is meaningful to compare the different variable selection schemes (lasso and ridge) with different shrinkage strength in other context data (business data in this paper).

### *4.4. Summary of the MCMC Estimation Algorithm*

We infer the parameters in the proposed RSD model via MCMC using the full conditional distributions described below. All the conditional distributions are in closed forms and hence can be easily sampled using a standard Gibbs sampler (see Section S1 in the Supplement Material for their detailed derivation). For each iteration of MCMC, we sample individual parameters using the following steps:

1) For each location $i = 1, \ldots, n$, draw the segment membership

   $g_i|\text{all others} \sim \text{Categorical}(\tilde{q}_{i1}, \ldots, \tilde{q}_{iK})$, where

   $\tilde{q}_{ig}|\text{all others} \propto q_g \frac{1}{\sigma_g/\sqrt{N_i}} \exp\left(-\frac{(y_i - \mathbf{x}_i \boldsymbol{\beta}_g)^2}{\sigma_g^2/N_i}\right) \cdot \sum_{h=1}^{M} p_{gh} \tau_g^2 \exp\left(-\tau_g^2 \|\mathbf{s}_i - \boldsymbol{\mu}_{gh}\|^2\right)$, $g = 1, \ldots, K$.

2) For each location $i = 1, \ldots, n$ draw the component membership

   $h_i|\text{all others} \sim \text{Categorical}(\tilde{p}_{i1}, \ldots, \tilde{p}_{iM})$,

   where $\tilde{p}_{ih}|\text{all others} \propto p_{g_ih} \tau_{g_i} \exp\left(-\tau_{g_i}^2 \|\mathbf{s}_i - \boldsymbol{\mu}_{g_ih}\|^2\right)$, $h = 1, \ldots, M$.

3) Sample the component mean $\boldsymbol{\mu}_{gh}|\text{all others} \sim \text{TBN}(\mathbf{m}_{gh}, v_g^2 I_2, -1, +1)$,

   with $\mathbf{m}_{gh} = \frac{\tau_g^2 \sum_{i=1}^{n} \mathbf{s}_i I(g_i=g) I(h_i=h)}{\tau_0^2 + \tau_g^2 \sum_{i=1}^{n} I(g_i=g) I(h_i=h)}$, $v_g^2 = \tau_0^2 + \tau_g^2 \sum_{i=1}^{n} I(g_i = g) I(h_i = h)$.

4) Sample the mixture component precision $\tau_g^2|\text{all others} \sim \text{Gamma}(\tilde{a}_{\tau,g}, \tilde{b}_{\tau,g})$, where $\tilde{a}_{\tau,g} = a_\tau +$

   $\sum_{i=1}^{n} I(g_i = g)$ and $\tilde{b}_{\tau,g} = b_\tau + \frac{\sum_{i=1}^{n}(\mathbf{s}_i - \boldsymbol{\mu}_{gh_i})^2 I(g_i=g)}{2}$.

5) Sample the regression coefficients

$$\boldsymbol{\beta}_g | \text{all others} \sim N\left(\left(\boldsymbol{X}_g^T \boldsymbol{W}_g \boldsymbol{X}_g + \boldsymbol{D}_g^{-1}\right)^{-1} \boldsymbol{X}_g^T \boldsymbol{W}_g \boldsymbol{Y}_g, \sigma_g^2 \left(\boldsymbol{X}_g^T \boldsymbol{W}_g \boldsymbol{X}_g + \boldsymbol{D}_g^{-1}\right)^{-1}\right),$$

where $\boldsymbol{D}_g = \text{diag}(\psi_{g1}, \dots, \psi_{gp})$.

6) Sample the error variance

$$\sigma_g^2 | \text{all others} \sim \text{InvGamma}(\tilde{a}_{\sigma,g}, \tilde{b}_{\sigma,g}),$$

where $\tilde{a}_{\sigma,g} = a_\sigma + \frac{n_g}{2}$, $\tilde{b}_{\sigma,g} = b_\sigma + \frac{\sum_{i=1}^n N_i(y_i - \mathbf{x}_i \boldsymbol{\beta}_g)^2 I(g_i = g)}{2}$.

6-1) For the ridge prior, simple set $\psi_{g1} = \dots = \psi_{gp} = 1/c = 100$.

6-2) For the lasso prior, sample

$$\frac{1}{\psi_{gj}} | \text{all others} \sim \text{InvGaussian}(\mu'_{gj}, \lambda^2), j = 1, \dots, p,$$

where $\mu'_{gj} = \sqrt{\frac{\lambda^2 \sigma_g^2}{\beta_{gj}^2}}$ and $\boldsymbol{D}_g = \text{diag}(\psi_{g1}, \dots, \psi_{gp})$.

## 4.5. Post-Processing

Once MCMC simulation has been conducted, we post-process the obtained sample to get the final clustering and parameter estimation results. Let $L$ be the total number of iterations in our MCMC sample. To estimate the segment memberships for individual observations $g_1, \dots, g_n$, we find the $l^*$th iteration of the MCMC sample which minimizes the following sum of squared differences:

$$l^* = \arg\min_l \sum_{i=1}^n \sum_{i=1}^n \left\{I(g_i^l = g_j^l) - d_{ij}\right\}^2,$$

where $d_{ij} = \sum_{l=1}^L I(g_i^l = g_j^l)/L$, and $g_i^l$ is the sampled segment membership for the $i^{\text{th}}$ observation in the $l^{\text{th}}$ iteration. This solution minimizes the posterior expected loss for estimated segment membership proposed by Binder (1978) in Bayesian model-based clustering (Dahl, 2006). Once the iteration for the optimal solution $l^*$ is determined, we rearrange the memberships $g_1, \dots, g_n$ for the $l^*$th iteration in a way that the resulting estimated memberships $g_1^*, g_2^*, \ldots, g_n^* \in \{1, \ldots, \widehat{K}\}$ where $\widehat{K}$ is the number of non-empty clusters for the $l^{*\text{th}}$ iteration. We then re-estimate the regression coefficients $\boldsymbol{\beta}_1, \ldots, \boldsymbol{\beta}_{\widehat{K}}$ and the error variances $\sigma_1^2, \ldots, \sigma_{\widehat{K}}^2$ for the non-empty clusters. We denote the estimated coefficients and the error variances in the post-processing stage by $\widehat{\boldsymbol{\beta}}_1, \ldots, \widehat{\boldsymbol{\beta}}_{\widehat{K}}$ and $\hat{\sigma}_1^2, \ldots, \hat{\sigma}_{\widehat{K}}^2$. For the model with ridge prior, we use the posterior mean $\widehat{\boldsymbol{\beta}}_g = \left(X_g^T W_g X_g + D_g^{-1}\right)^{-1} X_g^T W_g Y_g$ given in Equation (7) as the point estimator and also the estimated covariance $\hat{\sigma}_g^2 \left(X_g^T W_g X_g + D_g^{-1}\right)^{-1}$ to find the 95% credible interval. For the model with lasso prior, we use the weighted lasso estimates based on cross-validation implemented in *glmnet* package in R (Tibshirani et al., 2012) to estimate $\widehat{\boldsymbol{\beta}}_g$ for all non-empty $g = 1, \ldots, \widehat{K}$. While one can also consider using the Bayesian lasso estimator for $\boldsymbol{\beta}_g$, we choose to use a non-Bayesian estimator for its ability to provide a sparse solution with zero values.

## 5. A Monte Carlo Simulation Study

To test the performance of the proposed model, we conduct a Monte Carlo simulation study using multiple synthetic datasets that mimic various possible scenarios in the real applications. Many previous published studies have utilized this type of Monte Carlo analysis approach in testing the performance of various segmentation methods (e.g., Andrews, Ansari, and Currim 2002, DeSarbo and Cron 1988, Kim, Fong and DeSarbo 2012, Wedel and DeSarbo 1995). As benchmarks, we use two existing spatial segmentation models in Marketing, SAM by Hofstede et al. (2002) and SDS by Govind et al. (2018) discussed in Section 3.2. For SAM benchmark, since the original method does not account for label switching, we modified it by applying the post-processing approach described in Section 4.5. We also use the BIC instead of the Bayes Factors originally used by Hofstede et al. (2002) as our post-processing procedure chooses the segment labeling from one particular iteration in the MCMC chain as the 'best' estimate and hence the Bayes Factor is not applicable here. Our main objective in the simulation study is to investigate whether our RSD models provide better performances compared to the existing two benchmark approaches and how the two different regularizations for feature selection, the ridge and lasso affect the model performance across different conditions.

## 5.1. Study Design

We generate multiple simulated datasets by varying both the geographic characteristics and consumer behaviors. Two levels of all five factors are:

Factor 1) True number of clusters ($K^*$): $K^*=3$ vs. $K^*=6$;

Factor 2) Similarity of nearby neighboring locations: High similarity vs. Low similarity;

Factor 3) Density of spatial distributions of restaurants (i.e. overall number of observations $n$ within the fixed study domain): High density ($n=1{,}155$ within domain) vs. Low density ($n=563$ with domain) (See Section S2 in the Supplement Material for the details on how the sample size is determined.);

Factor 4) Number of predictors ($p$) that is covering the number of predictors in empirical application study: $p=4$ vs. $p=8$;

Factor 5) Proportion of active predictors (i.e. the proportion of nonzero coefficients $\beta_{g1}, \ldots, \beta_{gp}$ for each cluster $g$): 100% active predictors vs. 50% active predictors.

We define all two levels for each factor and hence generate $2^5 = 32$ simulated data sets. For generating the simulation datasets, we first generate the locations $s_1, \ldots, s_n$ and the true memberships $g_1^*, \ldots, g_n^*$ (see details in Section S2 in the Supplement Material). See Figure S1 (in the Supplement Material) for the spatial distribution of the restaurant locations and their assumed true memberships depending on different levels of the Factors 1 to 3. The total number of generated observations depends on the level of the density factor (Factor 3): $n = 1{,}155$ for the high density level and $n = 563$ for the low density level.

Then, we generate the true values of coefficients $\boldsymbol{\beta}_1^*, \ldots, \boldsymbol{\beta}_K^*$ that determine the relationship between $X_i$ and $Y_i$ in each segment. We set $p = 4$ or $8$ to vary the number predictors factor in Factor 4. We set the $p=4$ or $8$ considering to be much smaller than $n$, to reflect the fact that our real application has 4 predictor variables. We also set all or half of the coefficients to be nonzero depending on the level of active predictors factor (Factor 5). Each non-zero coefficient is independently generated from a uniform distribution with the range of [2, 15] and its sign is randomly determined with the probability of being positive equal to 0.5. Next, we generate the characteristics of the restaurants $(X_1, \ldots, X_n)$ and number of ratings that each of the restaurants has $(N_1, \ldots, N_n)$, which are independent of the segment memberships. Finally, we generate the response $Y_i$ based on the true memberships for the restaurants $g_1^*, \ldots, g_n^* \in \{1, \ldots, K^*\}$ and also the following model:

$$Y_i = X_i \boldsymbol{\beta}_{g_i^*}^* + \epsilon_i,$$

where the error $\epsilon_i \sim N\left(0, \frac{\sigma_0^2}{N_i}\right)$ with a pre-specified variance $\sigma_0^2 = 100$.

We summarize the varied factors in our simulation study in Table S1 in the Supplement Material. In each of the 32 simulated datasets, we have the training dataset $\{Y_i, X_i, N_i\}$ for $i = 1, \ldots, n$. We also generate 120 additional observations as a test data set that are evenly distributed over the $[0,1] \times [0,1]$ spatial domain in the same way described above and index them by $i = n + 1, \ldots, n + 120$. Using these datasets we compare the performance of our RSD model with the two benchmark methods, SAM and SDS. For each case, we first fit each model based on the training dataset, and then estimate the memberships for the test dataset $g_{n+1}, \ldots, g_{n+120}$ by copying the memberships of the nearest observations in the training dataset. We evaluate the model performance using the following four evaluation metrics:

i) Error in the estimated number of clusters, computed as $\text{DIFFK} = |\widehat{K} - K^*|$.

ii) Adjusted rand index (ARI, Rand, 1971, Hubert and Arabie 1985) for the test dataset, which evaluates the similarity between the true and estimated memberships (taking a value in [0,1], where higher the value the better the performance is).

iii) Root mean squared prediction errors (RMSPE) for the test dataset ($Y_i$), defined as

$$\text{RMSPE} = \sqrt{\sum_{i=n+1}^{n+120}(Y_i - X_i \widehat{\boldsymbol{\beta}}_{g_i})^2 / 120},$$

vi) Root mean squared error (RMSE) for the coefficients ($\beta_{g_i p}$), defined as

$$\text{RMSE} = \sqrt{\sum_{j=1}^{p} \sum_{i=n+1}^{n+120} (\beta_{g_i^* j}^* - \hat{\beta}_{g_i j})^2 / 120p}.$$

## *5.2. Results and Discussion*

Main objectives of the simulation study are (i) how using the proposed RSD improves model performances compared to the benchmarks and (ii) whether ridge or lasso regularization leads to a better results in terms of the above metrics. To highlight our answer for each question, we focus our discussion on how the performance of each model compares to that of RSD with ridge (RSD-ridge henceforth). To be more specific, for SAM, SDS and RSD with lasso (RSD-lasso henceforth) we compute the difference in the performance metric between each method and the RSD-ridge for all 32 cases and graphically examine the distributions of the differences.

The results are summarized as boxplots in Figure 1, where the results by RSD-ridge are used as a reference point. In the upper-left panel (a) for the differences in DIFFK, positive values indicate that the model has higher errors in estimating the number of segments compared to RSD-ridge. The result here indicates that RSD-ridge shows a better performance than the two benchmark models and also RSD-lasso. Interestingly, the result shows that SDS has much higher errors in estimating the number of segment memberships than other methods. SDS actually overestimates the number of segments in all 32 cases, yielding 2-7 more clusters than the assumed truth (see Table S1 in the Supplement Material). This is probably due to the mismatch between the likelihood function used for the actual model fitting and that for segment number selection in SDS. On the other hand, SAM always chooses $K^* = 3$, and hence always underestimates the number of latent segments. These results show that, at least under our simulation study setting, the existing methods can result in a notably inaccurate estimate for the true number of segments $K^*$.

In the upper-right panel (b) of Figure 1, negative values indicate that the model has lower values of ARI compared to the reference RSD-ridge, meaning that RSD-ridge outperforms that model in terms of estimating segment memberships. The plot here shows that RSD-ridge certainly outperforms the two benchmark models, SAM and SDS in terms of estimating the accurate segment memberships. However, the third boxplot in the panel (b) shows that RSD-ridge and RSD-lasso have almost no difference in ARI, meaning that the two methods have almost similar performances in segment membership recovery.

In the lower-left panel (c), positive values indicate that the model has higher RMSPE values and hence it is less accurate for predicting the response variable compared to the reference RSD-ridge. The plot shows that RSD-ridge has a better prediction performance than the two benchmark models but a similar prediction performance as RSD-lasso. The lower-right panel (d) is almost similar as the panel (c) except that the RMSE for estimating the regression coefficients is used as evaluation matrix. The plot shows that RSD-ridge outperforms the two benchmark models in terms of recovery of true segment coefficients. Moreover, it also slightly outperforms RSD-lasso, meaning that RSD-ridge is slightly more efficient than RSD-lasso in estimating the regression coefficients. In fact, this does make sense, as lasso penalization might impose unnecessarily strong shrinkage when $p \ll n$ (which is the case in our empirical application data) and hence lasso penalization underperforms than ridge penalization in the two levels of Factor 4.

[Insert Figure 1 about here]

One might argue that the two-levels of Factor 4 (p=4 or 8) might not fully examine potential benefits of strong shrinkage by lasso, and so we additionally conduct model comparison between RSD-ridge and RSD-lasso with high dimensionality data. With $K^*=6$ (Factor 1), the low similarity of nearby clusters (Factor 2), and the low density of spatial density (Factor 3), we increase the number of regression coefficients to either $p =30$ and 100. In both cases, we assume that only 10 predictor variables are active (i.e., moderately or highly sparse active regression coefficients). From Table 1, with the simulation data with large number of independent variables (p=30 or 100), RSD-ridge underestimates the true number of spatial segments, to the point that the method seems to be unable to capture any latent spatial structure in the simulated data (i.e., -3 for *p*=30 and -5 for *p*=100). The predicted responses and the estimated coefficients seem comparatively inaccurate. In contrast, RSD-lasso shows better performances in spatial clustering and also produces better estimations for true coefficients and better prediction for the true responses (see Table 1). Consistent with the regularization literature, when the number of features is moderately small (e.g., p=4 or 8 in this study), the ridge works slightly better than the lasso. However, as the number of features becomes large as p=30 or 100, (although *p* is not larger than *n*) the study proves the advantages of lasso regularization in the spatial mixture regression setting. As expected, we found that the performance gap between ridge and lasso becomes increased from p=30 to p=100.

[Insert Table 1 about Here]

In summary, through this Monte Carlo simulation study our RSD methods outperform the two traditional spatial market segmentation benchmarks in terms of all four evaluation metrics. As per the comparison between the ridge and lasso regularizations in RSD, we found that the ridge regularization shows a bit better performance than lasso penalization when the number of independent variables is moderately small (e.g., p=4). However, when the number of independent variables are large (e.g., p=100), we found the RSD with lasso penalization can have strong advantages over RSD with ridge. Thus, when the data includes many features (independent variables; large *p*), model users should consider lasso penalization. But, when the data includes small number of features (small *p*), model users should consider ridge penalization. This choice depending on the data type (i.e., large *p* vs. small *p*) can help business managers correctly estimate regression coefficients, which can lead more educated resource allocation across spatial segments (e.g., which feature(s) should be focused by manager for a certain regional segment).

## 6. Empirical Application: Restaurant Customer Satisfaction Study

### *6.1. Predictive Validation Comparison*

We now return to our main application problem, spatial analysis of restaurant customer satisfaction data from the Yelp. We first comparatively examine the predictive performance of four competing existing and proposed spatial segmentation models: SAM by Hofstede et al. (2002), SDS by Govind et al. (2018), and the two proposed models of RSD-ridge and RSD-lasso. We randomly hold out 10%, 20%, and 30% of the observational data as testing data set and fit the four competing models to the training data set. We then predict the observed consumer satisfactions with the test data set, and evaluate the prediction performance using Root Mean Square Prediction Error (RMSPE). We summarize the experiment results in Table 2. The results show that both RSD-ridge and RSD-lasso show smaller RMSPEs than traditional spatial segmentation methods of SAM and SDS across all three types of proportions. In specific, across 10%, 20% and 30% of test data set, the RSD-ridge produces somewhat stable 0.62, 0.631 and 0.636 of RMSPEs. Next, the RSD-lasso produces seemingly similar to the RSD-ridge but bit worse and unstable 0.782 for 10% case, 0.6304 for 20% case and 0.687 for 30% case. In comparison, two traditional methods of SAM and SDS provides clearly worse prediction performances such as 2.437 (10% case), 2.386 (20% case) and 2.374 (30% case) for SAM and 1.196 (10% case), 1.176 (20% case) and 1.195 (30% case) for SDS. It indicates that RSD models are better solutions in predictive ability for this dataset. Between two proposed approaches, RSD-ridge shows overall stable and better prediction performances (only except 20% test data case) compared to the RSD-lasso. This finding seems consistent with the simulation studies in previous section, given this application data has somewhat small number of four independent variables ($p=4$).

[Insert Table 2 about here]

## 6.2. Segment Profiles and Managerial Implications

Both proposed models of RSD-ridge and RSD-lasso are applied to the entire data set. The results by the both models are shown along with profiles of the automatically derived four segments. The various demographic information of the segments on town-level are obtained from the American Community Survey (ACS) conducted by the U.S. Census Bureau. To avoid any possible label switching issues (Jasra et al. 2005), we have re-arranged the segment labels for the RSD-lasso so that the same segment labels from the two models refer to similar spatial segments. Tables 3 and 4 provide the results of parameter estimates by the RSD-ridge and the RSD-lasso, respectively. The statistically significant coefficients found by the RSD-ridge are roughly similar to non-zero (selected) coefficients in the RSD-lasso, but there are some notable differences. Given stronger regularization scheme of lasso prior, the coefficients found significant by the RSD-ridge are often estimated as 0 or much smaller values by the RSD-lasso, which means the RSD-lasso provides more sparsity in variable selection. For examples, for the coefficients of Price, while the RSD-ridge provides significant coefficients across 3 Segments, the RSD-lasso provides 2 non-zero coefficients which seem smaller values than coefficients form the RSD-ridge. In addition, the coefficients for Category Spanning (CSPAN) variable are all estimated as 0 or smaller values by RSD-lasso while RSD-ridge has found that in some spatial segments the coefficients for CSPAN are significantly and also large enough to play an important role in determining the consumer satisfaction. In fact, CSPAN has been found to play a significant role in determining the consumer satisfaction in a previous study (Kovács and Johnson 2014) based on a similar dataset but segment-level analysis was not conducted before. These might indicate that the lasso regularization imposes overly strong regularization on the coefficients, possibly resulting in over-shrinkage towards zeros. Consistent with the Monte Carlo simulation study in the previous section, depending on the number of independent variables, model users may want to choose an appropriate type of regularization (i.e., lasso prior for very high dimensional data vs. ridge prior for somewhat low dimensional data). Given the ridge regularization is preferable in this case of four independent variables, we focus on the results by RSD-ridge from now.

[Insert Tables 3 and 4 about here]

The spatial segmentation results by the RSD-ridge are visualized in Figure 2. Many restaurants are clustered around the DC center area and other restaurants are located along with radial roads away from the center of DC. Also, we observe a spatial dependency pattern such that restaurants in near distances are grouped in one same segment, which is modeled in our spatial segmentation method. Restaurants in Segments 1 and 2 look mostly located around suburb areas, a bit away from the center area of Washington DC. For Segments 3 and 4, many restaurants of these segments tend to be clustered around the center area of Washington DC.

[Insert Figure 2 about here]

The results in Table 3 demonstrate that the four restaurant-level variables have varying influences on customer satisfaction across different segments. Combining the segmentation results and the ACS data, we found that Segment 1 is the most ethnically diverse with relatively high household income (second highest among the four segments). Consumers in Segment 1 show a stronger preference for non-chain restaurants than those in any other segment ($\beta = -0.41$) and have a preference for non-fast food restaurants as well ($\beta = -1.31$). Restaurants' category spanning, on the other hand, appears to have non-significant relationship with customer satisfaction in this ethnically diverse segment. These results imply that adopting a non-chain and non-fast food format can be advantageous for customer satisfaction for these ethnically diverse, high-income areas and that the restaurants that are located, or plan to be located, in the Segment 1 need to be cautious about their modes of operations.

Restaurants in Segment 2 are located at the areas with the lowest level of resident education with only 53.6% of the population aged over 25 having a bachelor's degree. Interestingly, the coefficient of CSPAN is negative and significant only in this segment, indicating that restaurants that span multiple categories reduce overall customer satisfaction in this segment. As we will find soon, in comparison, the coefficient of the CSPAN in Segment 4, which has the highest level of education, is positive and significant, implying that category spanning in the highly educated segment is rewarded in the form of customer satisfaction. These findings are consistent with research findings on category spanning and cultural sociology (Chae 2016; Ollivier 2008; Peterson and Kern 1996); organizations (restaurants) whose features cause them to be assigned to multiple categories (Kovács and Hannan 2010, p. 175) generally have lower chances of success because their spanning identities often confuse customers and are misunderstood, ignored, and devalued (Hsu, Hannan, and Koçak 2009; Wry, Lounsbury, and Jennings 2014). However, this trend flips among highly educated social elites, who have culturally omnivorous tastes and preferences for spanning. Thus, restaurants located in Segment 4 are more encouraged to span multiple market categories to appeal to educated cultural omnivores whereas restaurants located in Segment 2 are better off maintaining a solid and narrow market identity by staying in a single market category.

Segment 3 has the lowest income level with median household income of about $79,000. In this segment, fast food restaurants are positively related to customer satisfaction. This finding implies that, in contrast to the case of Segments 1 and 2, being a fast food restaurant is not penalized but rather appreciated by consumers in Segment 3. Finally, for Segment 4, it is also worth to note that, in addition to the significantly positive effect of category spanning (previously explained earlier), price had a negative and significant effect on customer satisfaction, which has the highest income and education level. Especially, the magnitude of the price effect ($\beta = -0.17$) is the highest in this segment compared with the other segments. This result implies that expectations towards expensive restaurants vary across customers with different levels of income and education and that affluent and educated elites have a higher standard for evaluating expensive restaurants, leading to miserly ratings for those establishments.

## 7. Summary and Discussion

We propose a spatial market segmentation method based on regularized Bayesian spatial mixture regression with Dirichlet process Gaussian mixture prior. First, our approach estimates the number of segments $K^*$ simultaneously along with other model parameters in a highly automated and statistically sound fashion. Therefore, the model automatically explores plausible choices for $K^*$ and selects the value that maximizes the posterior density without user's input. Moreover, this approach provides a more statistically sound way to determine the value of $K^*$ as the model used for determining the value of $\widehat{K}$ is consistent with the model used for estimating other parameters. In many of traditional approaches, the number of segments is heuristically determined by fitting the clustering model for pre-determined values of $K^*$ and then finding the value of $\widehat{K}$ that results in the smallest heuristics indicator (e.g., Bayes Factor or BIC). This sometimes requires manual exploration by a user, because the plausible range for $K^*$ has to be often determined by trial and error.

Second, the proposed method allows highly flexible spatial segmentation, where each segment can be represented as a collection of multiple components. The approach can even represent the situation where multiple geographically separate clusters exist within one segment; we argue that this is important for our application, because neighborhoods with similar demographics (e.g., household income, education level, etc.) are expected to show similar behavioral patterns even if they are not adjacent to each other. For example, two suburb regions with similar demographics should show similar consumer behavior even if they are located at the opposite sides of a big city (e.g., Washington DC in this paper).

Third, the sampling scheme of the proposed model provides a unified way of inference for the two different regularization methods, the ridge and lasso regularizations. In fact, the sampling scheme for the lasso regularization is almost identical to the ridge regularization except for the hyper-parameters for the prior variances $\psi_{g1}, \dots, \psi_{gp}$. So the proposed method can provide flexibility for employing the two different types of regularizations. Therefore, it is easy for business practitioners to design a software that allows a user to select one of the two regularization methods depending on the types of data (e.g., the number of covariates and sample size) and research purposes (e.g., ideal number of selected drivers for marketing resource allocation).

As limitations, our method does not assume spatial dependence for the error component in the regression model and hence possibly ignore the spatial effect that are not captured by the covariates considered in our model. Assuming spatial dependence however requires a careful modeling specification to avoid spatial confounding effects (e.g., Hodges and Reich, 2010, Huge and Haran 2012). Addressing these issues will provide an interesting avenue for future research, both in theory and application.

## Acknowledgements

We thank B. J. Reich for distributing his code for his version of spatial Dirichlet process Gaussian mixture model freely on the Web (https://www4.stat.ncsu.edu/~reich/code/SpatClust.R).

**Table 1. Model Comparison between Ridge and Lasso with High-dimensionality Simulation Data**

| N of IVs | DIFFK | | ARI | | RMSPE (Prediction) | | RMSE (Coefficients) | |
|---|---|---|---|---|---|---|---|---|
| $p$ | Ridge | lasso | ridge | lasso | ridge | lasso | ridge | lasso |
| 30 | -3 | 0 | 0.322 | 0.959 | 20.102 | 8.186 | 3.596 | 1.067 |
| 100 | -5 | 0 | 0 | 0.959 | 29.954 | 8.776 | 2.919 | 0.701 |

**Table 2. Prediction Performances for Test Data Measured by RMSPE**

| | SAM | | SDS | | The Proposed RSD-ridge | | The Proposed RSD-LASSO | |
|---|---|---|---|---|---|---|---|---|
| Test Data Set Proportion | N of clusters | RMSPE | N of clusters | RMSPE | N of clusters | RMSPE | N of clusters | RMSPE |
| 10% | 3 | 2.437 | 3 | 1.196 | 5 | 0.620 | 5 | 0.782 |
| 20% | 3 | 2.386 | 4 | 1.176 | 4 | 0.631 | 4 | 0.6304 |
| 30% | 3 | 2.374 | 5 | 1.195 | 4 | 0.636 | 4 | 0.687 |

**Table 3. Analysis Results by the Proposed RSD Regression with Ridge Prior**

| Segment | 1 | 2 | 3 | 4 |
|---|---|---|---|---|
| Intercept | 3.9738* | 4.2042* | 3.2691* | 3.9822* |
| Price | -0.1389* | -0.0481* | 0.0393 | -0.1712* |
| CSPAN | 0.0247 | -0.0368* | 0.0408* | 0.0643* |
| Chain | -0.4148* | -0.3570* | -0.1500* | -0.2112* |
| Fast Food | -1.3087* | -1.7268* | 0.2491* | 0.3058 |
| Ethnic Diversity | 0.6186 | 0.6027 | 0.5987 | 0.5353 |
| High Education | 0.5655 | 0.5360 | 0.5413 | 0.6623 |
| Adjusted Median Household Income | $97,636 | $82,296 | $79,113 | $104,030 |
| No. Of Obs. | 381 | 478 | 767 | 439 |
| Proportions | 18% | 23% | 37% | 21% |

*Notes:* * The 95% credible interval does not contain 0

**Table 4. Analysis Results by the Proposed RSD Regression with Lasso Prior**

| Segment | 1 | 2 | 3 | 4 |
|---|---|---|---|---|
| Intercept | 3.8589 | 4.0591 | 3.3713 | 3.9296 |
| Price | -0.0885 | 0 | 0 | -0.0978 |
| CSPAN | 0 | 0 | 0.0087 | 0.0063 |
| Chain | -0.252 | -0.2555 | -0.1399 | -0.1831 |
| Fast Food | -0.8267 | -1.5575 | 0 | 0.2941 |
| Ethnic Diversity | 0.5986 | 0.6067 | 0.6137 | 0.5184 |
| High Education | 0.5986 | 0.5319 | 0.5223 | 0.6701 |
| Adjusted Median Household Income | $97,705 | $78,272 | $74,329 | $11,6366 |
| No. Of Obs. | 435 | 420 | 792 | 418 |
| Proportions | 21% | 20% | 38% | 20% |

*Notes:* Non-zero coefficients are selected variables with Lasso regularization

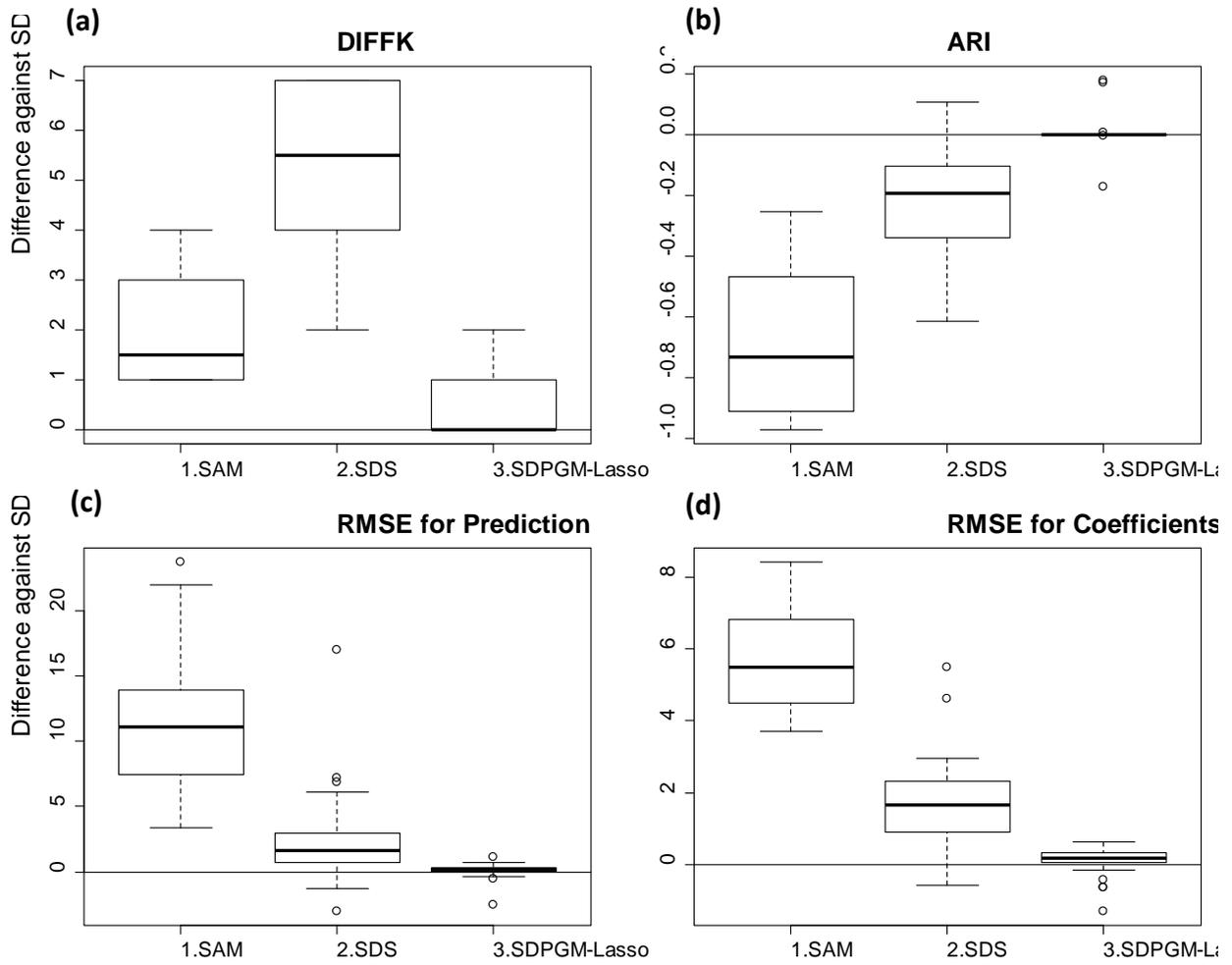

**Fig. 1. Model Performances against the Proposed RSD-Ridge that is used as Reference Point.**

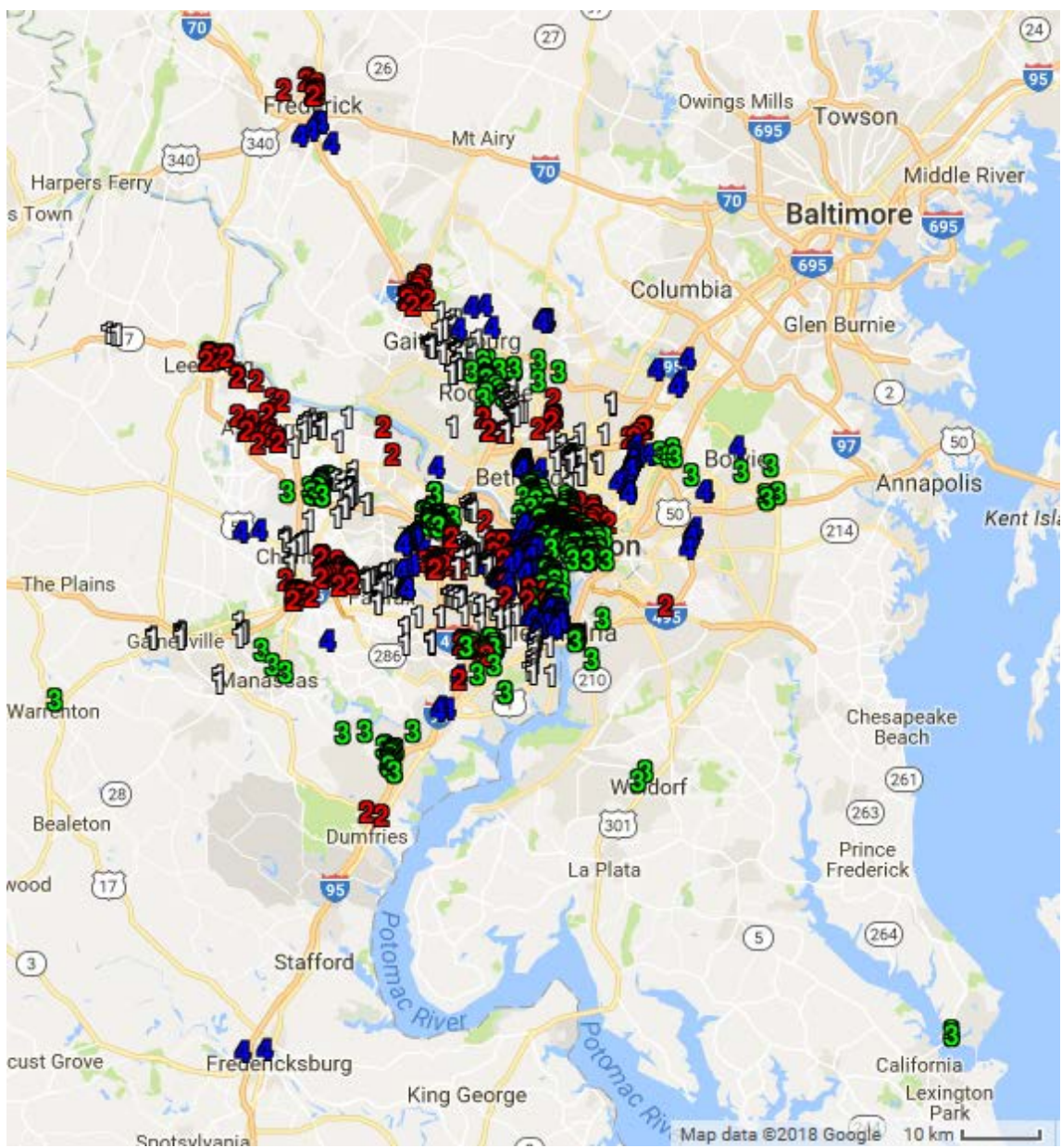

**Fig. 2. Segmentation Result by the RSD Regression with Ridge Prior**